\documentclass[12pt, a4paper] {article}

\usepackage{amsfonts}
\usepackage{verbatim}
 \usepackage{graphicx}

\begin{document}

\begin{center}
 
{\bf {\large Augmented Superfield Approach to Non-Yang-Mills Symmetries of Jackiw-Pi Model: Novel Observations}}

\vskip 1 cm

{\bf Saurabh Gupta$^{(a)}$, R. Kumar$^{(b)}$} \\
{{\it $^{(a)}$The Institute of Mathematical Sciences, \\CIT Campus, Chennai - 600 113, India\\
$^{(b)}$Physics Department, Center of Advanced Studies,} \\
{\it Banaras Hindu University, Varanasi - 221 005, India}} \\
{\small {\bf e-mails: saurabh@imsc.res.in, raviphynuc@gmail.com}}

\end{center}
\vskip 1 cm

\noindent
{\bf Abstract:} 
We derive the off-shell nilpotent and absolutely anticommuting Becchi-Rouet-Stora-Tyutin (BRST) as well as anti-BRST symmetry 
transformations corresponding to the non-Yang-Mills symmetry transformations of $(2 + 1)-$ dimensional Jackiw-Pi 
(JP) model within the framework of ``augmented'' superfield formalism. The Curci-Ferrari restriction, which is a hallmark 
of non-Abelian 1-form gauge theories, does not appear in this case.  
One of the novel features of our present investigation is the derivation of proper (anti-)BRST symmetry transformations 
corresponding to the auxiliary field $\rho$ that can not be derived by any conventional means.

\vskip 1.5 cm



\newpage

\noindent 
{\bf 1. Introduction}\\

\noindent 
The non-Abelian 1-form gauge theories are at the heart of the Standard Model (SM) of particle physics which accounts for the 
three out of four fundamental interactions of nature. The only missing link of SM, so far, is the existence of exotic Higgs 
boson which is responsible for the mass generation of bosons as well as fermions. However, the recent results in particle 
physics indicate the existences of a new boson. Whether this newly discovered boson is the Higgs boson of SM is 
not yet conclusive. 

In the view of above, other models for the mass generation have been considered in various dimensions of spacetime 
\cite{ djt1,djt2,ft,la1}. These models have generated a rejuvenate interest in this area of theoretical high energy physics.
It is worthwhile to mention, in particular, about the 4D topologically massive (non-)Abelian gauge theories \cite{la2,la3}  where
there is a merging of 1-form and 2-form fields through the celebrated topological $B \wedge F$ term. In these models the  1-form gauge 
field acquires a mass without taking any recourse to the Higgs mechanism. These topologically massive models have been throughly studies 
within the framework of superfield and Becchi-Rouet-Stora-Tyutin (BRST) formalism \cite{srm,rm1,rm2,mft}. 
The construction of a 4D consistent, renormalizable and unitary 
non-Abelian 2-form gauge theory is still an open problem, though some attempts have been made in this direction \cite{mka}.

Nevertheless, it is interesting to have a lower dimensional model which does not encounter such issues as of 4D topologically
massive models.  The Jackiw-Pi (JP) model in $(2 + 1)$-dimensions of spacetime is one such model \cite{jp}. In this
model, the gauge-invariance, mass and parity are respected simultaneously. Apart from the usual Yang-Mills (YM) symmetries this
model is also endowed with an another symmetry called non-Yang-Mills (NYM) symmetries. 

The Hamiltonian formulation and constraint analysis of JP model have been carried out \cite{dayi}, whereas BRST symmetries and 
Slovnov-Taylor identities, corresponding to YM symmetries, have also been established \cite{cima}. Recently, we have applied 
superfield formalism to derive the off-shell nilpotent and absolutely anticommuting  (anti-)BRST symmetry transformations of 
JP model corresponding to the usual YM symmetries \cite{sgjp}. One of the novel outcomes of this investigation is the 
derivation of the (anti-)BRST symmetry
transformations for the auxiliary field $\rho$ which is neither generated by the (anti-)BRST charges nor obtained by the 
requirement of nilpotency and/or absolute anticommutativity of the (anti-)BRST symmetries of 3D JP model. 

There are two equivalent ways to generalize a classical local continuous gauge symmetry to the quantum level, namely; (i) the BRST 
symmetry and, (ii) the anti-BRST symmetry. It is a well established fact that the anti-BRST symmetries do not play just a decorative 
part in the BRST formalism but have fundamental importance (see, e.g. \cite{oj,hw1,hw2} for details). Thus, keeping the above in mind, 
we have derived the proper BRST as well as anti-BRST symmetry transformations corresponding to the YM symmetries of the JP model by exploiting 
the ``augmented'' superfield approach to BRST formalism \cite{sgjp}. There we have purposely restricted ourselves only up to the
 YM symmetries of the JP model.

The prime motivations involved behind our present investigation are listed below. First and foremost, the derivation of full set of 
{\it proper} [i.e. off-shell nilpotent and absolutely anticommuting] (anti-)BRST symmetry transformations corresponding to the 
NYM gauge symmetries of JP model by exploiting the power and strength of ``augmented'' superfield approach to BRST formalism. 
Second, to examine the Curci-Ferrai (CF) condition, which is a hallmark of 1-form non-Abelian gauge theories, in 
the context of NYM symmetries of JP model. Finally, to obtain the appropriate Lagrangian density which respects the (anti-)BRST symmetry 
transformations derived with the help of ``augmented'' superfield formalism.

Our present paper is organized in the following manner. In section 2, we discuss the gauge symmetries associated with the JP model. 
Our section 3 deals with the derivation of the off-shell nilpotent and absolutely anticommuting (anti-)BRST symmetry transformations 
corresponding to the NYM gauge symmetries of the JP model within the framework of ``augmented'' superfield formalism. Our section 4 
contains the derivation of the (anti-)BRST invariant Lagrangian density and section 
5 deals with the conserved (anti-)BRST currents (and their corresponding charges). The ghost symmetry transformations and the 
algebra satisfied by all the generators are incorporated in section 6. Finally, in section 7, we summarize our key results and point 
out some future directions.  

In Appendix A, we deal with the nilpotency of the (anti-)BRST symmetry transformations (and their corresponding generators) 
within the framework of superfield formalism. 

{\it Conventions and notations:} We adopt here the notations and conventions such that the flat Minkowski metric in 3D is 
$\eta_{\mu\nu} = diag (-1, +1, +1)$ and 3D Levi-Civita tensor follows $\varepsilon_{\mu\nu\eta} \varepsilon^{\mu\nu\eta} = - 3!, \; 
\varepsilon_{\mu\nu\eta} \varepsilon^{\mu\nu\sigma} = - 2! \; \delta^\sigma_\eta,$ etc., with $\varepsilon_{012} = +1 = 
- \varepsilon^{012}$. Here Greek indices $\mu,\nu,\eta,... = 0,1,2$. The dot product, cross product and covariant derivative are defined 
as: $A \cdot B = A^a B^a, \; A \times B = f^{abc} A^a B^b T^c, \; D_\mu C^a = \partial_\mu C^a - g \;(A_\mu \times C)^a $ in the $SU(N)$ 
Lie algebraic space spanned by the generators $T^a$. In the above, $g$ is a coupling constant and the $SU(N)$ generators follow the 
algebra: $[T^a, T^b] = f^{abc} T^c$ with $a,b,c... = 1,2,3,..., N^2-1$ where $f^{abc}$ is totally antisymmetric structure constant 
\cite{wein}. \\

\noindent
{\bf 2. Preliminaries: (Non-)Yang-Mills symmetries}\\
 
\noindent
We begin with the following Lagrangian density of $(2 + 1)-$ dimensional $(3D)$ Jackiw-Pi (JP) model \cite{jp}
\begin{eqnarray}
{\cal L}_0 &=& - \frac{1}{4} F^{\mu\nu} \cdot F_{\mu\nu} - \frac{1}{4} \big(G^{\mu\nu} 
+ g\; F^{\mu\nu} \times \rho\big) \cdot \big(G_{\mu\nu} + g \;F_{\mu\nu} \times \rho\big) \nonumber\\
&+& \frac {m}{2} \; \varepsilon^{\mu\nu\eta} \; F_{\mu\nu} \cdot \phi_\eta, \label{jp1}
\end{eqnarray}
where $F_{\mu\nu} = \partial_\mu A_\nu - \partial_\nu A_\mu - g \;(A_\mu \times A_\nu)$ is 2-form $[F^{(2)} =  \frac {1}{2!} 
(dx^\mu \wedge dx^\nu) F_{\mu\nu} \cdot T]$ curvature tensor corresponding to the 1-form $(A^{(1)} = dx^\mu A_\mu \cdot T)$
field $A_\mu$ and $G_{\mu\nu} = D_\mu \phi_\nu - D_\nu \phi_\mu$ is an another 2-form [$G^{(2)} = \frac {1}{2!} (dx^\mu \wedge dx^\nu) 
G_{\mu\nu} \cdot T$] field strength tensor corresponding to the 1-form $(\phi^{(1)} = dx^\mu \phi_\mu \cdot T)$ field $\phi_\mu$. 
The 2-form curvature tensor $F^{(2)}$ has its origin in Maurer-Cartan equation $F^{(2)} = d A^{(1)} + i g \;(A^{(1)} \wedge A^{(1)})$ 
whereas $G^{(2)}$ is obtained from $G^{(2)} = d \phi^{(1)} + i g \;[\phi^{(1)} \wedge A^{(1)} +   A^{(1)} \wedge \phi^{(1)}]$. 
The above mentioned 1-form fields (i.e. $A_\mu$ and $\phi_\mu$)  have opposite parity which makes this a parity conserving model. 
Furthermore, $\rho$  is a scalar field and $m$ represents the mass parameter.

The above Lagrangian density (\ref{jp1}) respects two sets of local symmetry transformations, the usual Yang-Mills (YM) 
gauge transformations ($\delta_1$) and non-Yang-Mills (NYM) gauge transformations ($\delta_2$), namely \cite{cima, sgjp};
\begin{eqnarray}
&&\delta_1 A_\mu = D_\mu \Lambda, \qquad \delta_1 \phi_\mu = - g\;(\phi_\mu \times \Lambda), 
\qquad \delta_1 \rho = - g\;(\rho \times \Lambda), \nonumber\\
&&\delta_1 F_{\mu\nu} = - g\;(F_{\mu\nu} \times \Lambda), \qquad \delta_1 G_{\mu\nu} 
= - g\;(G_{\mu\nu} \times \Lambda),
\end{eqnarray}
\begin{eqnarray}
\delta_2 A_\mu = 0, \qquad \delta_2 \phi_\mu = D_\mu \Omega  , \qquad \delta_2 \rho 
= +\; \Omega , \qquad \delta_2 F_{\mu\nu} = 0,
\end{eqnarray}
where $\Lambda = \Lambda \cdot T$ and $\Omega = \Omega \cdot T$ are $SU(N)$ valued infinitesimal gauge parameters corresponding 
to YM and NYM gauge transformations, respectively. It is straightforward to check that the following is true:
\begin{eqnarray}
\delta_1 {\cal L}_0 = 0, \hskip 2.cm \delta_2 {\cal L}_0 = \partial_\mu \Big[\frac {m}{2}\; 
\varepsilon^{\mu\nu\eta} \;F_{\nu\eta} \cdot \Omega \Big].
\end{eqnarray} 
This means that $\delta_1$ and $\delta_2$ are the symmetry transformations of the Lagrangian density (\ref{jp1}).
At this juncture, we would like to mention that all {\it proper} (anti-)BRST symmetry transformations corresponding to the usual 
YM symmetry transformations $(\delta_1)$ has already been derived in our earlier endeavor \cite{sgjp}. Here, we shall focus  
on NYM symmetry transformations ($\delta_2$). \\

\noindent
{\bf 3. (Anti-)BRST symmetries: Augmented superfield formalism}\\

\noindent
We apply augmented superfield formalism \cite{lbt1,lbt2} to derive the off-shell nilpotent and absolutely anticommuting (anti-)BRST symmetry 
transformations corresponding to the NYM symmetries of the JP model. For this purpose, first, we generalize the $3$D bosonic fields 
[$A_\mu (x), \phi_\mu (x), \rho(x)$] and fermionic (anti-)ghost [$(\bar \beta) \beta(x) $] fields of the theory to their corresponding 
superfields on $(3,2)-$dimensional supermanifold characterized by the variables $x^M$ $(M = \mu, \theta, \bar \theta)$ where
$\mu (= 0, 1, 2)$ stands for the usual spacetime variables and $\theta, \bar\theta$ are Grassmannian variables (with 
$\theta^2 = \bar\theta^2 = 0, \; \theta \bar\theta + \bar\theta \theta = 0$). 
Then we expand these superfields along the Grassmannian directions. It is worthwhile to mention here that the superfield 
expansion of $A_\mu (x)$ along the Grassmannian directions is equal to $A_\mu (x)$ itself because of the fact $\delta_2 A_\mu = 0$. 
The above statement can also be, mathematically, incorporated as 
\begin{equation}
 \tilde A_\mu (x,\theta,\bar\theta) \; = \; A_\mu (x),
\end{equation}
here $\tilde A_\mu (x,\theta,\bar\theta) $ is the superfield corresponding to the field $A_\mu (x)$ when latter is generalized onto the 
$(3, 2)-$dimensional supermanifold. Rest of the superfields\footnote{The superfields corresponding to $\phi_\mu(x),$ $\rho(x),$ 
$\beta(x)$ and $\bar\beta(x)$ are $\tilde\phi_\mu (x, \theta, \bar\theta),$ $ \tilde \rho (x, \theta, \bar\theta),$ 
$ \tilde \beta (x, \theta, \bar\theta), $ $\tilde {\bar \beta} (x, \theta, \bar\theta) $, respectively, when the formers are 
generalized onto the $(3, 2)$-dimensional supermanifold parametrized by the variables $x^M$ (with $M = \mu, \theta, \bar\theta$). } of 
the present theory can be expanded along the Grassmannian directions ($\theta, \bar \theta$) as follows
\begin{eqnarray}   
\tilde \phi_\mu (x,\theta,\bar\theta) &=& \phi_\mu (x) + \theta\; \bar R_\mu (x) + \bar\theta\;
R_\mu (x) + i \;\theta \;\bar\theta\; S_\mu (x),\nonumber\\
\tilde \rho (x,\theta,\bar\theta) &=& \rho (x) + \theta \; \bar b (x)
+ \bar\theta\; b (x) + i \;\theta \;\bar\theta \;q(x), \nonumber\\
\tilde \beta (x,\theta,\bar\theta) &=& \beta (x) + i\;\theta\; \bar R_1 (x) + i\; \bar\theta\;
R_1 (x) + i \;\theta \;\bar\theta\; s (x),\nonumber\\
\tilde {\bar \beta }(x,\theta,\bar\theta) &=& \bar \beta (x) + i\;\theta\; \bar R_2 (x) + i\;\bar\theta\;
R_2 (x) + i \;\theta \;\bar\theta\; \bar s (x), \label{expa}
\end{eqnarray}  
where $b (x), \bar b(x), R_\mu (x), \bar R_\mu (x), s (x), \bar s (x)$ are fermionic secondary fields. The rest of the  
secondary fields, i.e. $q (x), R_1 (x), R_2 (x), \bar R_1 (x), \bar R_2 (x), S_\mu (x)$, are bosonic in nature. 

Second, we suitably choose a physical quantity (in some sense) and demand that this quantity must  remain unaffected by the presense 
of Grassmannian variables when the former is generalized onto the $(3,2)-$dimensional supermanifold. In this connection, it is worthwhile to 
note that $[G_{\mu\nu} + g \; (F_{\mu\nu} \times \rho)]$ remains invariant under the NYM gauge transformations $(\delta_2)$, i.e.
\begin{equation}
 \delta_2 \; [G_{\mu\nu} + g \; (F_{\mu\nu} \times \rho)] \; = \; 0.
\end{equation}
Thus, this combination serves our purpose. Therefore, we have the following gauge invariant 
restriction (GIR)
\begin{equation}
 \tilde G_{MN} + g \; (\tilde F_{MN} \times \tilde \rho) \; = \; G_{\mu\nu} + g \; (F_{\mu\nu} \times \rho), \label{gir}
\end{equation}
where $\tilde G_{MN} $ is super curvature tensor that can be derived from $\tilde G^{(2)} = \tilde d \tilde\phi^{(1)} 
+ i g \; [ (\tilde \phi^{(1)} \wedge \tilde A^{(1)}) +  (\tilde A^{(1)} \wedge \tilde \phi^{(1)}) ] \equiv \frac{1}{2!} 
(dZ^M \wedge dZ^N) \tilde G_{MN}$ and $\tilde F_{MN}$ is super field strength tensor having its origin in 
$\tilde F^{(2)} = \tilde d \tilde A^{(1)} + i g \; (\tilde A^{(1)} \wedge \tilde A^{(1)}) \equiv \frac{1}{2!} 
(dZ^M \wedge dZ^N) \tilde F_{MN}$. 
Here $\tilde d$ denotes the super exterior derivative and $\tilde A^{(1)}$  is super 1-form connection which are the generalizations 
of the exterior derivative $d$ and 1-form connection $A^{(1)}$, respectively. Furthermore, $\tilde \phi^{(1)}$ and $\tilde \rho$ 
are the superfields corresponding to the 1-form field $\phi^{(1)}$  and  scalar field $\rho$, respectively. These quantities can be,
mathematically, summarized as follows (when generalized onto the $(3,2)$-dimensions of spacetime),  
\begin{eqnarray}
d \to \tilde d = dZ^M \partial_M &\equiv& dx^{\mu} \; \partial_{\mu} + d \theta \; \partial_{\theta} + d {\bar \theta} \; \partial_{\bar \theta}, \nonumber\\
\phi^{(1)} \to \tilde\phi^{(1)} = dZ^M \tilde \phi_M &\equiv& dx^\mu \tilde \phi_\mu (x,\theta,\bar\theta) 
+ d \theta \tilde {\bar \beta} (x,\theta,\bar\theta) + d \bar \theta \tilde \beta (x,\theta,\bar\theta),\nonumber\\
A^{(1)} \to \tilde A^{(1)} = dZ^M \tilde A_M &\equiv&  dx^\mu \tilde A_\mu (x, \theta, \bar\theta) = dx^\mu A_\mu (x), \nonumber\\
\rho \to \tilde \rho &=&  \tilde \rho(x, \theta, \bar \theta). \label{sup}
\end{eqnarray}
It is worthwhile to mention that the above GIR [cf. (\ref{gir})] can also be written in the following fashion 
\begin{eqnarray}
& \tilde d \tilde\phi^{(1)} + i g  \Big[ (\tilde \phi^{(1)} \wedge \tilde A^{(1)}) +  (\tilde A^{(1)} \wedge \tilde \phi^{(1)}) 
-  (\tilde F^{(2)} \wedge \tilde \rho)  + (\tilde \rho \wedge \tilde F^{(2)}) \Big] & \nonumber\\ 
& = d \phi^{(1)} + i g \Big[ (\phi^{(1)} \wedge A^{(1)}) +  (A^{(1)} \wedge \phi^{(1)}) - (F^{(2)} \wedge \rho) 
 +  (\rho \wedge F^{(2)}) \Big].  &\label{gir2}
\end{eqnarray}
Third, we substitute for the super exterior derivative and superfields from (\ref{sup}) into the l.h.s. of (\ref{gir2}) and similarly, 
substituting for the exterior derivative and 3D fields in r.h.s., too. 
Then comparing the coefficients of corresponding wedge products from both the sides, we have following expressions 
\begin{eqnarray}
&& D_\mu \tilde \phi_\nu - D_\nu \tilde \phi_\mu + g (F_{\mu\nu} \times \tilde \rho)
= D_\mu \phi_\nu - D_\nu  \phi_\mu + g (F_{\mu\nu} \times \rho), \nonumber\\
&& \partial_\mu \tilde {\bar \beta} - \partial_\theta \tilde \phi_\mu 
- g (A_\mu \times \tilde {\bar \beta})  = 0, \quad  \partial_\theta \tilde \beta + \partial_{\bar \theta} 
\tilde {\bar \beta} = 0, \nonumber\\
&&  \partial_\mu \tilde \beta - \partial_{\bar \theta} \tilde \phi_\mu 
- g (A_\mu \times \tilde \beta) = 0, \qquad \partial_\theta \tilde {\bar \beta} = 0, \qquad
\partial_{\bar \theta} \tilde  \beta = 0. \label{11}
\end{eqnarray}  
Finally, we substitute for the superfields expansion from (\ref{expa}) in the above expressions as listed in (\ref{11}). Thus, we obtain following relationships amongst the basic, auxiliary 
and secondary fields of the theory 
\begin{eqnarray}
&&R_\mu = D_\mu \beta, \qquad \bar R_\mu = D_\mu \bar \beta, \qquad S_\mu  =  D_\mu R, \nonumber\\
&&b = \beta, \quad \bar b =  \bar \beta, \quad q = R, \quad R_1 = \bar R_2 = s = \bar s = 0. \label{rel}
\end{eqnarray}
Here we have made the choice $R_2 = - \bar R_1 = R$. Substituting, the above relationships (\ref{rel}) into the super-expansion 
of the superfields [cf. (\ref{expa})], we obtain the following explicit expansions:
\begin{eqnarray}
\tilde \phi_\mu ^{(g)} (x, \theta, \bar\theta) &=& \phi_\mu (x) + \theta \; [D_\mu \bar \beta (x)] + \bar \theta \; [D_\mu \beta (x)] 
+ \theta \bar \theta \; [i D_\mu R (x)] \nonumber\\
&\equiv& \phi_\mu (x) + \theta \; [s_{ab} \; \phi_\mu (x)] + \bar \theta \; [s_{b} \; \phi_\mu  (x)] 
+ \theta \bar \theta \; [s_b s_{ab} \; \phi_\mu (x)], \nonumber\\ 
\tilde \rho^{(g)} (x, \theta, \bar \theta) &=& \rho (x) + \theta \; [\bar\beta (x)] + \bar \theta \; [\beta (x)] 
+ \theta \bar\theta \; [i R (x)] \nonumber\\
&\equiv& \rho (x) + \theta \; [s_{ab} \; \rho (x)] + \bar \theta \; [s_b \; \rho (x)] 
+ \theta \bar \theta \; [s_b s_{ab} \; \rho (x)], \nonumber\\
\tilde \beta^{(g)} (x, \theta, \bar\theta) &=& \beta (x) - \theta \; [i R (x)] \nonumber\\
&\equiv& \beta (x) + \theta\; [s_{ab} \; \beta (x)], \nonumber\\
\tilde {\bar \beta}^{(g)} (x, \theta, \bar\theta) &=& \bar \beta (x) + \bar \theta \; [i R (x)] \nonumber\\
&\equiv& \bar \beta (x) + \bar \theta [s_b \; \bar \beta (x)], \label{g}
\end{eqnarray}
where the superscript $(g)$ on the superfields refers to the superexpansions of the superfields obtained after the application of 
GIR [cf. (\ref{gir})]. Thus, we can easily read the (anti-)BRST symmetry transformations ($s_{a(b)}$) from the above equations. 
These transformations are listed below
\begin{eqnarray}  
&& s_b \phi_\mu = D_\mu \beta, \qquad s_b \bar \beta = iR, \qquad s_b \rho = \beta,\nonumber\\
&& s_b G_{\mu\nu} = - g(F_{\mu\nu} \times \beta), \qquad s_b[F_{\mu\nu}, \;A_\mu, \; \beta,\; R] = 0, \label{sb1}
\end{eqnarray} 
\begin{eqnarray}
&& s_{ab} \phi_\mu = D_\mu \bar \beta, \qquad s_{ab}  \beta = - iR, \qquad s_{ab} \rho = \bar \beta,\nonumber\\
&& s_{ab} G_{\mu\nu} = - g(F_{\mu\nu} \times \bar \beta ), \qquad s_{ab}[F_{\mu\nu}, \;A_\mu, \; \bar \beta,\; R] = 0. \label{sb2}
\end{eqnarray}
It is worthwhile to mention that the equations in (\ref{g}) imply that: $s_b \longleftrightarrow \displaystyle \lim_{\theta \to 0} 
\; (\partial/ \partial \bar \theta), \; 
s_{ab} \longleftrightarrow \displaystyle \lim_{\bar\theta \to 0} \; (\partial/ \partial \theta)$. Thus, the (anti-)BRST symmetry 
transformations are related with the translational generators along the Grassmannian directions of the $(3,2)$-dimensional 
supermanifold by the above mentioned mappings. 
These (anti-)BRST symmetry transformatons, that are obtained from the augumented superfield formalsim, are nilpotent of 
order two (i.e. $s_{a(b)}^2 = 0$) and absolutely anticommuting in nature. This is true because of the fact that $(\partial_\theta)^2 = 
(\partial_{\bar \theta})^2 = 0$ and $\partial_\theta \; \partial_{\bar \theta} + \partial_{\bar\theta} \; \partial_\theta = 0$. 
We also capture nilpotency and absolute anticommutativity of the (anti-)BRST symmetries within the framework of superfield formalism in our 
Appendix A.

One of the crucial findings of the augmented superfield formalism applied to the NYM case of JP model is the non-existence of 
Curci-Ferrai (CF) restriction. This restriction is the hallmark of non-Abelian 1-form gauge theories and plays a central role 
for the proof of absolute anticommutativity of the (anti-)BRST symmetry transformations. The above condition appears naturally 
within the framework of superfield formalism when the latter is applied to the YM case of JP model \cite{sgjp}. 
But, in the present case the CF restriction does not exist.\\

\noindent
{\bf 4. (Anti-)BRST invariant Lagrangian density}\\

\noindent 
The most appropriate expression for the (anti-)BRST invariant Lagrangian density (corresponding to the NYM symmetry of JP model) 
can be obtained in the following manner: 
\begin{eqnarray}
{\cal L}_b &=& {\cal L}_0 + s_b \; s_{ab} \left[\frac {i}{2}\;\phi_\mu \cdot \phi^\mu 
+ \frac {1}{2}\; \beta \cdot \bar \beta \right]\nonumber\\
&\equiv& {\cal L}_0 - s_{ab} \; s_{b} \left[\frac {i}{2}\;\phi_\mu \cdot \phi^\mu 
+ \frac {1}{2}\; \beta \cdot \bar \beta \right]. 
\end{eqnarray}
The terms in the square brackets are Lorentz scalars and they are chosen in such a fashion that the mass dimension
and ghost number of each term is one and zero, respectively. The above Lagrangian density, in its full blaze of glory,
can be written as follows:
\begin{eqnarray}
{\cal L}_b &=& - \frac{1}{4} F^{\mu\nu} \cdot F_{\mu\nu} - \frac{1}{4} \big(G^{\mu\nu} 
+ g\; F^{\mu\nu} \times \rho\big) \cdot \big(G_{\mu\nu} + g \;F_{\mu\nu} \times \rho\big) \nonumber\\
&+& \frac {m}{2} \varepsilon^{\mu\nu\eta} \; F_{\mu\nu} \cdot \phi_\eta  + \frac {1}{2}\; R \cdot R + R \cdot (D_\mu \phi^\mu) 
- i\;(D_\mu \bar \beta)\cdot(D^\mu \beta). \label{lag}
\end{eqnarray}
It is interesting to note that in the present case (NYM) no gauge-fixing and Faddeev-Popov ghost terms are required for
the gauge field $A_\mu$ in the above Lagrangian density. The reason, behind this observation, is that the field $A_\mu$ does not
transform under (anti-)BRST symmetry transformations i.e. $s_{(a)b} \; A_\mu = 0$ [cf. (\ref {sb1}), (\ref{sb2})]. Therefore, 
we do not have any gauge-fixing and Faddeev-Popov ghost terms corresponding to the gauge field $A_\mu$. 

The above Lagrangian density ${\cal L}_b$ remains quasi-invariant under the (anti-)BRST symmetry transformations. This 
can be checked as follows:
\begin{eqnarray} 
 s_b \; {\cal L}_b &=& \partial_\mu \left[\frac {m}{2}\; \varepsilon^{\mu\nu\eta} \;\beta \cdot F_{\nu\eta} 
+ R \cdot D^\mu \beta \right],\nonumber\\
s_{ab} \; {\cal L}_b &=& \partial_\mu \left[\frac {m}{2}\; \varepsilon^{\mu\nu\eta} \;\bar\beta \cdot F_{\nu\eta} 
+ R \cdot D^\mu \bar \beta \right]. \label{inv}
\end{eqnarray}
Thus, the corresponding actions (i.e. $\int d^3x \; {\cal L}_b $) remains invariant under the (anti-)BRST symmetry transformations
due to the validity of Gauss's divergence theorem for the physically well-defined fields. \\

\noindent
{\bf 5. Conserved charges: Novel observations} \\

\noindent
The action corresponding to the Lagrangian density ${\cal L}_b$ remains invariant [cf. (\ref{inv})] under the continuous (anti-)BRST 
symmetry transformations [cf. (\ref{sb1}), (\ref{sb2})]. This implies, according to Noether's theorem, the existence of conserved 
currents (and corresponding conserved charges). Thus, exploiting the basics of Noether's theorem, following (anti-)BRST currents 
$J^\mu_{(a)b}$,   
\begin{eqnarray}
J^\mu_b &=& - (D_\nu \beta)\cdot \left[G^{\mu\nu} + g (F^{\mu\nu} \times \rho) \right]
+ R \cdot (D^\mu \beta) - \frac {m}{2}\; \varepsilon^{\mu\nu\eta} \;\beta \cdot F_{\nu\eta}, \nonumber\\ 
J^\mu_{ab} &=& - (D_\nu \bar \beta)\cdot \left[G^{\mu\nu} + g (F^{\mu\nu} \times \rho) \right]
+ R \cdot (D^\mu \bar \beta) - \frac {m}{2}\; \varepsilon^{\mu\nu\eta} \;\bar \beta \cdot F_{\nu\eta}, \label{jmu}
\end{eqnarray}
can be derived from the Lagrangian density (\ref{lag}). The conservation law (i.e. $\partial_\mu J^\mu_{(a)b} = 0 $) can 
be proven by using following Euler-Lagrange (E-L) equations of motion 
\begin{eqnarray}
&& D_\mu F^{\mu\nu} - g\;D_\mu \left[\left(G^{\mu\nu} + g F^{\mu\nu} \times \rho\right)\times \rho\right]
 + g\; \left[\left(G^{\mu\nu} + g F^{\mu\nu} \times \rho \right)\times \phi_\mu \right] \nonumber\\
&& + \; m \;\varepsilon^{\mu\eta\nu}\; D_\mu \phi_\eta + g (R \times \phi^\nu) - ig \;(\bar \beta \times D^\nu \beta)  
+ ig \;(\beta \times D^\nu \bar \beta) = 0, \nonumber\\
&& D_\mu (G^{\mu\nu} + g F^{\mu\nu} \times \rho) + \frac {m}{2}\; \varepsilon^{\mu\eta\nu} \;F_{\mu\eta} - D^\nu R = 0, \nonumber\\
&& R + D_\mu \phi^\mu = 0, \quad D_\mu (D^\mu \bar \beta) = 0, \quad D_\mu (D^\mu  \beta) = 0.         
\end{eqnarray}
These Euler-Lagrange equations of motion have been derived with the help of Lagrangian density (\ref{lag}).
Furthermore, we can re-express the conserved (anti-)BRST currents [cf. (\ref{jmu})], using above E-L equations of motion,
in the following cute and convenient form:   
\begin{eqnarray}
J^\mu_b = R \cdot \big(D^\mu \beta\big) - \beta \cdot \big(D^\mu R\big) - \partial_\nu [ \beta \cdot ( G^{\mu\nu} + g F^{\mu\nu} 
\times \rho)  ], \nonumber\\  
J^\mu_{ab} = R \cdot \big(D^\mu \bar \beta\big) - \bar \beta \cdot \big(D^\mu R\big) - \partial_\nu [ \bar \beta \cdot 
( G^{\mu\nu} + g F^{\mu\nu} \times \rho) ].  
\end{eqnarray}
The zeroth component of above conserved currents (i.e. $\int d^2 x \; J^0_{(a)b} $) is defined as the    
conserved (anti-)BRST charges $Q_{(a)b}$, namely;
\begin{eqnarray}
Q_b &=& \int d^2x \, \Big[R \cdot \big(D^0 \beta\big) - \beta \cdot \big(D^0 R\big)\Big], \nonumber\\  
Q_{ab} &=& \int d^2x \, \Big[R \cdot \big(D^0 \bar \beta\big) - \bar \beta \cdot \big(D^0 R\big)\Big]. \label{qb}
\end{eqnarray}
It is interesting to point out that the above mentioned (anti-)BRST charges can be re-expressed 
in the following manner:
\begin{eqnarray}
 Q_b &=& \int d^2x\, s_b \, \Big[R(x) \cdot  \phi^0 (x) - i\, \beta (x) \cdot \big(D^0 \, \bar \beta \big)(x)
 \Big]\nonumber\\
&\equiv& i\, \int d^2x\, s_b\, s_{ab}\, \Big[\beta (x) \cdot \phi^0 (x) \Big], \nonumber\\
Q_{ab} &=& \int d^2x\, s_{ab} \, \Big[R(x) \cdot  \phi^0 (x) + i\, \bar \beta (x) \cdot \big(D^0 \, \beta \big)(x)
 \Big]\nonumber\\
&\equiv& - i\, \int d^2x\, s_{ab}\, s_{b}\, \Big[ \bar \beta (x) \cdot \phi^0 (x) \Big], 
\end{eqnarray}
where $s_{(a)b}$ are the (anti-)BRST symmetry transformations given in (\ref{sb2}) and (\ref{sb1}), respectively.
In this form, the nilpotency and absolute anticommutativity properties of (anti-)BRST charges can be checked in 
a straightforward manner because of the fact:  $s_{(a)b}^2 = 0$ and $s_b s_{ab} + s_{ab} s_b = 0.$

These charges $Q_{(a)b}$ turn out to be the generators of the (anti-)BRST symmetry transformations (\ref{sb2}) and (\ref{sb1}), 
respectively. It can be explicitly checked  from the following relationship 
\begin{equation}
 s_r \; \Phi = \pm \; i \; [\Phi, Q_r]_\pm,  \quad  (r = b, ab), \label{gen}
\end{equation}
where $\Phi$ is generic field of the theory. The  $(\pm)$ signs in the subscript of  square bracket 
stand for the (anti)commutator for the field $\Phi$ being (fermionic) bosonic in nature whereas the  $(\pm)$
signs in front of the square bracket have been chosen judiciously (see for detail \cite{srcan}). It is interesting to note that the following 
algebraic structure 
\begin{eqnarray}
&& s_b Q_b = - i\;\big \{Q_b, \; Q_b\big\}=  0, \quad
s_{ab} Q_{ab} = - i\;\big \{Q_{ab}, \; Q_{ab}\big\}=  0, \nonumber\\
&& s_b Q_{ab} = - i\;\big \{Q_{ab}, \; Q_b\big\}=  0, \quad 
s_{ab} Q_b = - i\;\big \{Q_b, \; Q_{ab}\big\} =  0,   
\end{eqnarray}
is derived from the transformations (\ref{sb1}) and (\ref{sb2}) when we exploit the expressions $Q_{(a)b}$ from (\ref{qb}) and use 
the definition of generator from (\ref{gen}). 

It is worthwhile to mention, at this juncture, that the conserved [i.e. $\dot Q_{(a)b} = 0 $] and nilpotent [i.e. $Q^2_{(a)b} = 0$] 
(anti-)BRST charges are incapable of generating the (anti-)BRST symmetry transformations for the auxiliary field $`\rho$' (though they 
generate (anti-)BRST symmetry transformations for all the basic fields of the theory). Moreover, the absolute anticommutativity and/or
nilpotency properties of the (anti-)BRST symmetry transformations are also inadequate to generate the (anti-)BRST symmetry transformations 
for the auxiliary field $\rho$. This is one of the novel features of this theory. \\

\noindent
{\bf 6. Ghost symmetry and BRST algebra} \\

\noindent
The Lagrangian density (\ref{lag}) respect the following continuous global $[\Lambda \neq \Lambda (x)]$ scale symmetry transformations, 
namely;
\begin{equation}
\beta \to e^{+ \Lambda} \; \beta, \qquad \bar \beta \to e^{- \Lambda} \; \bar \beta, \qquad
\Psi  \to e^{0} \; \Psi, \quad(\Psi = A_\mu, \phi_\mu, \rho, R).
\end{equation}
The $(\pm)$ signs, in the exponentials, stand for the ghost numbers of the corresponding (anti-)ghost fields. It is evident that 
$\beta$  and $\bar \beta$ have the ghost number $(+1)$ and $(-1)$, respectively, whereas the ghost number for rest of the fields 
(i.e. $A_\mu, \phi_\mu, \rho, R$) is equal to zero. The infinitesimal version  of the above global scale 
transformations ($s_g$) is given as:
\begin{eqnarray}
s_g \; \beta = + \Lambda \; \beta, \qquad s_g \; \bar \beta = - \Lambda \; \bar \beta, 
\qquad s_g \; [A_\mu, \phi_\mu, \rho, R] = 0. \label{gh}
\end{eqnarray}
These are the symmetry transformations of the Lagrangian density (\ref{lag}) because $s_g \; {\cal L}_b = 0$. Exploiting the 
Noether's theorem, in the context of above ghost symmetry transformations ($s_g$), we obtain following conserved current 
\begin{equation}
J^\mu_g \; = \; i\;\left[\beta \cdot D^\mu \bar \beta +  \bar \beta \cdot D^\mu \beta\right]. 
\end{equation}
The conservation law ($\partial_\mu J^\mu_g = 0 $) can be easily proven. The temporal component of the above current 
leads to the conserved $(\dot Q_g = 0)$ ghost charge $Q_g$ as defined below:
\begin{equation}
Q_g = \int d^2 x \; J^0_g \; = \; i\int d^2 x \left[\beta \cdot D^0\bar \beta +  \bar \beta \cdot D^0 \beta\right].
\end{equation}
It turns out that the above ghost charge is the generator of the  
infinitesimal ghost-scale transformations (\ref{gh}). For instance, it can be checked that
$s_g \bar \beta = + i \Lambda \, [\bar \beta,\; Q_g] = - \Lambda\; \bar \beta.$
The (anti-)BRST charges [$Q_{(a)b}$] and the ghost charge ($Q_g$) follow the standard BRST algebra, namely;
\begin{eqnarray}
&&  Q^2_b = 0, \qquad Q^2_{ab} = 0, \qquad i \; [Q_g, Q_b] = Q_b, \qquad i\; [Q_{g}, Q_{ab}] = - Q_{ab}, \nonumber\\
&& \{Q_b, Q_{ab} \} \; = \;  Q_b Q_{ab} + Q_{ab} Q_b \; = \; 0, \quad [Q_g, Q_g] \; = \; 0, \quad Q_g^2 \neq 0. \label{alg}
\end{eqnarray}
Let $p$ be the ghost number of a state $|\psi\rangle_n$ (in the quantum Hilbert space of states), which is defined as follows:
\begin{eqnarray}
i\; Q_g \; |\psi\rangle_n \; = \; p \; |\psi\rangle_n.
\end{eqnarray} 
With the help of above algebra (\ref{alg}), it is
straightforward to check that the following relations are true, namely;
\begin{eqnarray}
i \,Q_g\, Q_b \,|\psi\rangle_n &=& (p + 1)\,Q_b \,|\psi\rangle_n,\nonumber\\
i \,Q_g\, Q_{ab}\, |\psi\rangle_n &=& (p - 1)\,Q_{ab} \,|\psi\rangle_n.
\end{eqnarray}
Thus, we conclude that the BRST charge $Q_b$ increases the ghost number by one whereas the anti-BRST charge $Q_{ab}$ decreases 
the same by one unit. In other words,  $Q_{(a)b}$ carry the ghost numbers $(\mp1)$, respectively.
These observations also reflect from the expressions of the (anti-)BRST and 
ghost charges where the ghost numbers of the fields are concerned.   \\

\noindent
{\bf 7. Conclusions} \\

\noindent
In our present endeavor, we have utilized the non-Yang Mills symmetries of the 3D JP model at the classical level and generalized it to the 
quantum level [i.e. the (anti-)BRST symmetry transformations]. In fact, we have derived the full set of off-shell nilpotent and 
absolutely anticommuting  (anti-)BRST symmetry transformations corresponding to the NYM symmetries of the JP model within the 
framework of ``augmented'' superfield formalism. Until now, there was no conventional derivation available (in the known literature) for the 
(anti-)BRST symmetry transformations of the JP model. 

Furthermore, the derivation of {\it proper} (anti-)BRST symmetry transformations for the auxiliary field $\rho$  is a crucial 
finding of present investigation. This is  because of the fact that the (anti-)BRST symmetry transformations corresponding to $\rho$ can 
neither be derived from the conserved (anti-)BRST charges nor by the requirement of 
nilpotency and/or  absolute anticommutativity property of the (anti-)BRST symmetry transformations. 

The non-existence of the Curci-Ferrai restriction, in the case of NYM symmetries of  3D JP model,  is yet another novel
observation of the present study. This is on the contrary to the YM case of JP model where the CF restriction emerges 
naturally within the framework of superfield formalism \cite{sgjp} and the absolute anticommutativity of the (anti-)BRST
symmetries is ensured by this CF restriction. 

Moreover, we have obtained the proper Lagrangian density which respects the above (anti-)BRST symmetry transformations. Apart from 
the usual (anti-)BRST symmetries this Lagrangian density is also endowed with one more continuous symmetry in the ghost sector (i.e. the
ghost symmetry). We have exploited this ghost symmetry to derive the ghost charge and finally, we have shown the standard BRST 
algebra that is obeyed by all the conserved charges of the theory.     

It would be a nice endeavor to take the combination of YM and NYM symmetry {\it together} and derive proper (anti-)BRST symmetry 
transformations and appropriate coupled Lagrangian densities corresponding to the combined symmetry. At present, these issues are
under investigation and our results will be reported in our future publications \cite{sgjp3}.  \\

\noindent

\noindent
{\bf Acknowledgements} \\

\noindent
We are grateful to R. P. Malik for enormous helpful suggestions and comments. One of us (RK) would 
like to thankfully acknowledge the financial support from UGC, New Delhi, Government of India.

\noindent\\
{\bf Appendix A}

\noindent\\
We capture the nilpotency and absolute anticommutativity properties of the (anti-)BRST symmetry transformations  
within the framework of superfield formalism. The above mentioned properties can be easily proven with the help of 
translational generators (i.e. $\partial_\theta, \partial_{\bar\theta} $) along the Grassmannian directions of the 
supermanifold. The nilpotency of the (anti-)BRST symmetry transformations is captured in the following 
manner:
\begin{eqnarray}
&& s_b \; \Longleftrightarrow \; \lim_{\theta \to 0} \frac{\partial}{\partial \bar\theta}, \qquad s_b^2 = 0 \Longleftrightarrow 
\Big(\frac{\partial} {\partial \bar \theta} \Big)^2 \; = \; 0, \nonumber\\
&& s_{ab} \; \Longleftrightarrow \; \lim_{\bar \theta \to 0} \frac{\partial}{\partial \theta}, \qquad s_{ab}^2 = 0 
\Longleftrightarrow  \Big(\frac{\partial} {\partial \theta} \Big)^2 \; = \; 0,
\end{eqnarray}
whereas the absolute anticommutativity property of the (anti-)BRST symmetry transformations are encoded in the following expression:
\begin{equation}
 s_b s_{ab} + s_{ab} s_b  =  0 \; \Longleftrightarrow\;  \frac{\partial}{\partial \theta} \frac{\partial}{\partial \bar \theta} + 
\frac{\partial}{\partial \bar \theta} \frac{\partial}{\partial \theta} =  0.
\end{equation}
In order to capture the nilpotency of the BRST charge $Q_{b}$ [cf. (\ref{qb})], within 
superfield formalism, we first  express $Q_{b}$  in terms of superfields as 
\begin{eqnarray}
&& Q_b = \lim_{\theta \to 0}\; \frac{\partial}{\partial \bar \theta} \int d^2x
\Big[R(x) \cdot  \tilde \phi^{0(g)} (x, \theta, \bar \theta) 
- i\,  \tilde \beta^{(g)} (x, \theta, \bar \theta) \cdot \big(D^0 \,\tilde {\bar \beta}^{(g)} \big)
(x, \theta, \bar \theta) \Big]\nonumber\\
&&\equiv \int d^2 x \int d \bar \theta \Big[R(x) \cdot  \tilde \phi^{0(g)} (x, \theta, \bar \theta) 
- i\,  \tilde \beta^{(g)} (x, \theta, \bar \theta) \cdot \big(D^0 \,\tilde {\bar \beta}^{(g)} \big)
(x, \theta, \bar \theta) \Big]. \label{qbs}
\end{eqnarray}
As a consequence, from the above expressions, it is clear that 
\begin{equation}
 \lim_{\theta \to 0} \frac{\partial}{\partial \bar \theta} \; Q_b = 0 \; \Longrightarrow \; Q_b^2 = 0,
\end{equation}
because of the fact that $\partial_{\bar \theta}^2 = 0$. The above equation precisely proves the nilpotency of the BRST charge $Q_b$.
It is interesting to note that the BRST charge [cf. (\ref{qbs})] can be re-expressed, in a compact way, as follows
\begin{equation}
 Q_b \; = \;  i\,\frac{\partial}{\partial \theta}\, \frac{\partial}{\partial \bar \theta} \int d^2x
\Big[\tilde \beta^{(g)} (x, \theta, \bar \theta) \cdot \tilde \phi^{0(g)} (x, \theta, \bar \theta) \Big].
\end{equation}
The proof of nilpotency is rather straightforward in the above mentioned form of $Q_b$ (because $\partial^2_\theta = \partial^2_{\bar\theta}
= 0$). Similarly, we can express the anti-BRST charges $Q_{ab}$ in the following fashion
\begin{eqnarray}
Q_{ab} &=& \lim_{\bar\theta \to 0}\; \frac{\partial}{\partial \theta} \int d^2x
\Big[R(x) \cdot  \tilde \phi^{0(g)} (x, \theta, \bar \theta) + i\, \tilde {\bar \beta}^{(g)} (x, \theta, \bar \theta) \cdot 
\big(D^0 \,\tilde {\beta}^{(g)} \big) (x, \theta, \bar \theta) \Big]\nonumber\\
&\equiv& \int d^2 x \int d \theta \Big[R(x) \cdot  \tilde \phi^{0(g)} (x, \theta, \bar \theta) + i\, \tilde {\bar \beta}^{(g)} (x, \theta, \bar \theta) \cdot 
\big(D^0 \,\tilde {\beta}^{(g)} \big) (x, \theta, \bar \theta) \Big] \nonumber\\
&\equiv&  i\,\frac{\partial}{\partial  \theta}\, \frac{\partial}{\partial \bar \theta} \int d^2x
\Big[\tilde {\bar \beta}^{(g)} (x, \theta, \bar \theta) \cdot \tilde \phi^{0(g)} (x, \theta, \bar \theta) \Big].
\end{eqnarray}
The proof of nilpotency of the anti-BRST charge rely on the nilpotency property ($\partial_\theta^2 = 0 $) of the Grassmannian 
derivative $\partial_\theta$, as given below:  
\begin{equation}
 \lim_{\bar\theta \to 0} \frac{\partial}{\partial \theta} \; Q_{ab} = 0 \; \Longrightarrow \; Q_{ab}^2 = 0. 
\end{equation}
Thus, we note that the nilpotency of the (anti-)BRST charges is encoded in $\partial^2_\theta = \partial^2_{\bar\theta} = 0$, when 
the latter is expressed in terms of the Grassmannian derivatives on $(3,2)$-dimensional supermanifold.  \\

\end{document}